\documentclass[preprint2]{aastex}
\usepackage{graphicx}
\usepackage{gensymb}
\usepackage{lineno}
\newcommand{\U}[1]{\ensuremath{\mathrm{\ #1}}}
\newcommand{\UU}[2]{\ensuremath{\mathrm{\ #1^{#2}}}}
\slugcomment{Submitted to ApJ Letters}

\shorttitle{Optical SETI of KIC 8462852}
\shortauthors{The VERITAS Collaboration}

\begin{document}


\title{A Search for Brief Optical Flashes Associated with the SETI Target KIC 8462852}


%\author{The VERITAS Collaboration \altaffilmark{1}}
%\affil{VERITAS affiliations}

\author{
A.~U.~Abeysekara\altaffilmark{1},
S.~Archambault\altaffilmark{2},
A.~Archer\altaffilmark{3},
W.~Benbow\altaffilmark{4},
R.~Bird\altaffilmark{5},
M.~Buchovecky\altaffilmark{6},
J.~H.~Buckley\altaffilmark{3},
K.~Byrum\altaffilmark{7},
J.~V~Cardenzana\altaffilmark{8},
M.~Cerruti\altaffilmark{4},
X.~Chen\altaffilmark{9,10},
J.~L.~Christiansen\altaffilmark{11},
L.~Ciupik\altaffilmark{12},
W.~Cui\altaffilmark{13},
H.~J.~Dickinson\altaffilmark{8,B},
J.~D.~Eisch\altaffilmark{8},
M.~Errando\altaffilmark{14},
A.~Falcone\altaffilmark{15},
D.~J.~Fegan\altaffilmark{5},
Q.~Feng\altaffilmark{13},
J.~P.~Finley\altaffilmark{13},
H.~Fleischhack\altaffilmark{10},
P.~Fortin\altaffilmark{4},
L.~Fortson\altaffilmark{16},
A.~Furniss\altaffilmark{17},
G.~H.~Gillanders\altaffilmark{18},
S.~Griffin\altaffilmark{2},
J.~Grube\altaffilmark{12},
G.~Gyuk\altaffilmark{12},
M.~H{\"u}tten\altaffilmark{10},
N.~H{\aa}kansson\altaffilmark{9},
D.~Hanna\altaffilmark{2},
J.~Holder\altaffilmark{19,28,A},
T.~B.~Humensky\altaffilmark{20},
C.~A.~Johnson\altaffilmark{21},
P.~Kaaret\altaffilmark{22},
P.~Kar\altaffilmark{1},
N.~Kelley-Hoskins\altaffilmark{10},
M.~Kertzman\altaffilmark{23},
D.~Kieda\altaffilmark{1},
M.~Krause\altaffilmark{10},
F.~Krennrich\altaffilmark{8},
S.~Kumar\altaffilmark{19},
M.~J.~Lang\altaffilmark{18},
T.~T.Y.~Lin\altaffilmark{2},
G.~Maier\altaffilmark{10},
S.~McArthur\altaffilmark{13},
A.~McCann\altaffilmark{2},
K.~Meagher\altaffilmark{24},
P.~Moriarty\altaffilmark{18},
R.~Mukherjee\altaffilmark{14},
D.~Nieto\altaffilmark{20},
S.~O'Brien\altaffilmark{5},
A.~O'Faol\'{a}in de Bhr\'{o}ithe\altaffilmark{10},
R.~A.~Ong\altaffilmark{6},
A.~N.~Otte\altaffilmark{24},
N.~Park\altaffilmark{25},
J.~S.~Perkins\altaffilmark{26},
A.~Petrashyk\altaffilmark{20},
M.~Pohl\altaffilmark{9,10},
A.~Popkow\altaffilmark{6},
E.~Pueschel\altaffilmark{5},
J.~Quinn\altaffilmark{5},
K.~Ragan\altaffilmark{2},
G.~Ratliff\altaffilmark{12},
P.~T.~Reynolds\altaffilmark{27},
G.~T.~Richards\altaffilmark{24},
E.~Roache\altaffilmark{4},
M.~Santander\altaffilmark{14},
G.~H.~Sembroski\altaffilmark{13},
K.~Shahinyan\altaffilmark{16},
D.~Staszak\altaffilmark{2},
I.~Telezhinsky\altaffilmark{9,10},
J.~V.~Tucci\altaffilmark{13},
J.~Tyler\altaffilmark{2},
S.~Vincent\altaffilmark{10},
S.~P.~Wakely\altaffilmark{25},
O.~M.~Weiner\altaffilmark{20},
A.~Weinstein\altaffilmark{8},
D.~A.~Williams\altaffilmark{21},
B.~Zitzer\altaffilmark{7}
}

\altaffiltext{A}{{\tt jholder@physics.udel.edu}}
\altaffiltext{B}{{\tt hughd@iastate.edu}}
\altaffiltext{1}{Department of Physics and Astronomy, University of Utah, Salt Lake City, UT 84112, USA}
\altaffiltext{2}{Physics Department, McGill University, Montreal, QC H3A 2T8, Canada}
\altaffiltext{3}{Department of Physics, Washington University, St. Louis, MO 63130, USA}
\altaffiltext{4}{Fred Lawrence Whipple Observatory, Harvard-Smithsonian Center for Astrophysics, Amado, AZ 85645, USA}
\altaffiltext{5}{School of Physics, University College Dublin, Belfield, Dublin 4, Ireland}
\altaffiltext{6}{Department of Physics and Astronomy, University of California, Los Angeles, CA 90095, USA}
\altaffiltext{7}{Argonne National Laboratory, 9700 S. Cass Avenue, Argonne, IL 60439, USA}
\altaffiltext{8}{Department of Physics and Astronomy, Iowa State University, Ames, IA 50011, USA}
\altaffiltext{9}{Institute of Physics and Astronomy, University of Potsdam, 14476 Potsdam-Golm, Germany}
\altaffiltext{10}{DESY, Platanenallee 6, 15738 Zeuthen, Germany}
\altaffiltext{11}{Physics Department, California Polytechnic State University, San Luis Obispo, CA 94307, USA}
\altaffiltext{12}{Astronomy Department, Adler Planetarium and Astronomy Museum, Chicago, IL 60605, USA}
\altaffiltext{13}{Department of Physics and Astronomy, Purdue University, West Lafayette, IN 47907, USA}
\altaffiltext{14}{Department of Physics and Astronomy, Barnard College, Columbia University, NY 10027, USA}
\altaffiltext{15}{Department of Astronomy and Astrophysics, 525 Davey Lab, Pennsylvania State University, University Park, PA 16802, USA}
\altaffiltext{16}{School of Physics and Astronomy, University of Minnesota, Minneapolis, MN 55455, USA}
\altaffiltext{17}{Department of Physics, California State University - East Bay, Hayward, CA 94542, USA}
\altaffiltext{18}{School of Physics, National University of Ireland Galway, University Road, Galway, Ireland}
\altaffiltext{19}{Department of Physics and Astronomy and the Bartol Research Institute, University of Delaware, Newark, DE 19716, USA}
\altaffiltext{20}{Physics Department, Columbia University, New York, NY 10027, USA}
\altaffiltext{21}{Santa Cruz Institute for Particle Physics and Department of Physics, University of California, Santa Cruz, CA 95064, USA}
\altaffiltext{22}{Department of Physics and Astronomy, University of Iowa, Van Allen Hall, Iowa City, IA 52242, USA}
\altaffiltext{23}{Department of Physics and Astronomy, DePauw University, Greencastle, IN 46135-0037, USA}
\altaffiltext{24}{School of Physics and Center for Relativistic Astrophysics, Georgia Institute of Technology, 837 State Street NW, Atlanta, GA 30332-0430}
\altaffiltext{25}{Enrico Fermi Institute, University of Chicago, Chicago, IL 60637, USA}
\altaffiltext{26}{N.A.S.A./Goddard Space-Flight Center, Code 661, Greenbelt, MD 20771, USA}
\altaffiltext{27}{Department of Physical Sciences, Cork Institute of Technology, Bishopstown, Cork, Ireland}
\altaffiltext{28}{Department of Physics and Space Science, Florida Institute of Technology, W. Melbourne, FL 32901, USA}


\begin{abstract}
  The F-type star KIC~8462852 has recently been identified as an
  exceptional target for SETI (search for extraterrestrial
  intelligence) observations. We describe an analysis methodology for
  optical SETI, which we have used to analyse nine hours of
  serendipitous archival observations of KIC~8462852 made with the
  VERITAS gamma-ray observatory between 2009 and 2015. No evidence of
  pulsed optical beacons, above a pulse intensity at the
  Earth of approximately $1\U{photon}\UU{m}{-2}$, is found. We also discuss the potential
  use of imaging atmospheric Cherenkov telescope arrays in searching
  for extremely short duration optical transients in general.
\end{abstract}


\keywords{extraterrestrial intelligence --- astrobiology --- stars:
  individual (KIC~8462852) --- techniques: photometric --- methods: observational}



\section{Introduction}

Over the course of its four-year primary mission, NASA's
\textit{Kepler} spacecraft provided photometric measurements of over
150,000 stars, sampled typically every 30 minutes. A number of the
resulting high-precision lightcurves exhibit unusual variability
patterns, which can generally be explained as the result of analysis
artifacts or by known astrophysical mechanisms.
\citet{2015arXiv150903622B} recently provided an in-depth study of
KIC~8462852 (TYC~3162-665-1), a star whose lightcurve was flagged as
unusual by members of the Zooniverse citizen science Planet Hunters
project \citep{2012MNRAS.419.2900F}. They identify the star as a main
sequence F3 V/IV star, and describe unique, aperiodic dips in the
stellar flux of up to 20\%, lasting for between 5 and 80
days. Archival photographic plates also show unprecedented
century-long dimming, at an average rate of $0.165\pm0.013$ magnitudes
per century \citep{2016arXiv160103256S}.  One possible explanation is
that the observations may be explained by the passage of a family of
exocomet fragments resulting from a single break-up event
\citep{2015arXiv150903622B,
  2015arXiv151108821B}. \citet{2016ApJ...816...17W} offer an
alternative ``extraordinary hypothesis'' that the lightcurve is
consistent with the existence of a collection of planet-sized
structures, or swarms of many smaller objects, placed in orbit by an
extraterrestrial civilization. They describe KIC~8462852 as an
outstanding target for SETI (search for extraterrestrial intelligence)
obsdervations. Radio frequency observations between 1 and 10 GHz using
the Allen Telescope Array from 15 October to 30 October, 2015, for
approximately 12 hours each day, did not find any evidence of a signal
\citep{2015arXiv151101606H}.

Astronomical SETI observations are most commonly conducted at radio
frequencies, close to the $21\U{cm}$ hydrogen line. However,
\citet{Schwartz61} noted over half a century ago that optical light
sources are also effective beacons, detectable over
interstellar distances. One promising method is to search for intense
pulses of optical or near-infrared photons from candidate star
systems: \citet{2004ApJ...613.1270H} have calculated that, using
current technology (10m reflectors as the transmitting and receiving
apertures and a $3.7\U{MJ}$ pulsed laser source), a $3\U{ns}$ optical
pulse could be produced that would be easily detectable at a distance
of $1000\U{ly}$, outshining starlight from the host system by a factor
of $10^4$. A number of dedicated searches for such signals have been
conducted or are under development (e.g. \cite{2004ApJ...613.1270H,
  2005AsBio...5..604S, 2007AcAau..61...78H,
  2014SPIEWright}). \citet{2015arXiv151202388S} performed a search for
periodic optical pulses from KIC~8462852 between October 29 and
November 9, 2015, using the $0.5\U{m}$ telescope of the Boquete
Optical SETI Observatory, with null results.  Alternative approaches
to optical SETI include searching for the spectral signatures of laser
emission either in the form of extremely narrow emission lines
\citep{2015PASP..127..540T} or as periodic signatures in the spectra
\citep{2012AJ....144..181B}.

The detection of nanosecond optical pulses from the night sky requires
large-aperture mirrors instrumented with fast photon
detectors. Ground-based gamma-ray telescopes have identical
requirements and can be used to search for SETI-like signals
\citep{Covault01, Eichler01, 2005ICRC, 2005neeu,
  2009AsBio...9..345H}. The requirement for
coincident signals between multiple independent telescopes, combined
with the ability to form a crude image of the light flash, allows
imaging atmospheric Cherenkov telescopes (IACTs) to perform such
searches parasitically during regular observations, with effectively
no background. We report here on the results of observations of
KIC~8462852, recorded serendipitously by the VERITAS gamma-ray
observatory.

\begin{figure*}
 \includegraphics[width=0.72\textwidth]{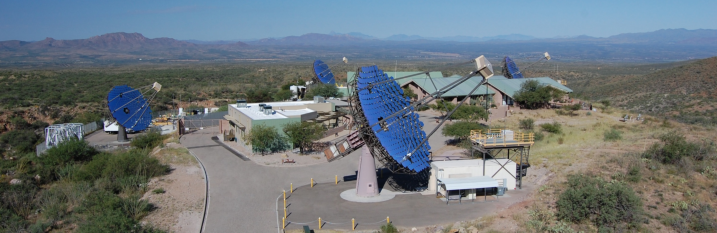}\hspace{0.1cm}\includegraphics[width=0.259\textwidth]{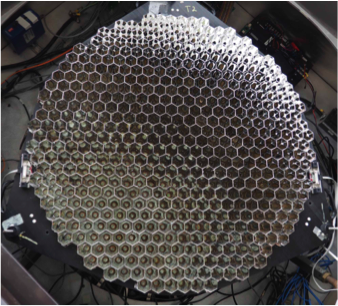}
 \caption{The VERITAS array in Arizona (left) and the PMT camera of a single
   telescope (right), that covers a total field-of-view
   of diameter $3.5\degree$. The hexagonal reflective light cones reduce the
   dead space between the circular PMTs.}\label{array}
\end{figure*}

\section{VERITAS Instrument and Observations}

VERITAS \citep{2002APh....17..221W} is an array of four IACTs located
at the Fred Lawrence Whipple Observatory in southern Arizona
(Fig.~\ref{array}).  The telescope optics follow a Davies-Cotton
design \citep{Davies57}, with a $12\U{m}$ aperture reflector and a
$12\U{m}$ focal length. The reflector comprises $345$ hexagonal mirror
facets, giving a total mirror area of $110\UU{m}{2}$. The telescopes
are mounted on steerable alt-azimuth positioners, and are arranged in
an approximate diamond formation with sides of roughly $100\U{m}$
length. At the focus of each telescope is a close-packed array of 499
photomultiplier tubes (PMTs; Fig.~\ref{array}). The PMT angular
spacing is $0.15\degree$, giving a roughly circular field-of-view for
each telescope of diameter $3.5\degree$. Dead space between the PMTs
is removed by the addition of reflecting hybrid-Winston cones to the
PMT front faces, with a hexagonal shape at the entrance and a circular
exit window \citep{2008ICRC....3.1437N}. The photodetectors (initially
Photonis XP2970 PMTs, upgraded to Hamamatsu R10560 Super Bialkali
PMTs in 2012) are sensitive throughout the visible wavelength range,
with a peak detection efficiency around $400\U{nm}$.

Ground-based gamma-ray astronomy is an indirect technique, which works
by forming an image of cascades of relativistic particles in the
Earth's atmosphere (``air showers'') using the few-ns pulse of
optical Cherenkov light that they generate. The telescopes therefore
do not integrate light continuously; rather, the PMT signals are
split, with one copy passed to a fast multi-level trigger system. A
telescope trigger is generated when the signals in at least three
adjacent PMTs cross a discriminator threshold within approximately
$5\U{ns}$. An array trigger is generated when at least two of the four
telescopes trigger within a coincidence window of $50\U{ns}$. On
receipt of a successful array trigger, all of the PMT signals are
read out using flash analog-digital converters (FADCs) which sample
the signal every $2\U{ns}$. The trigger rate is typically a few
hundred Hz, almost all of which is due to cosmic ray initiated
particle cascades. The VERITAS telescope design is described in more
detail in \citet{2006APh....25..391H}, and the IACT technique is
summarized in \citet{2015arXiv151005675H}.

%By measuring the
%intensity and morphology of the image in multiple telescopes, the
%nature of the cascade progenitor (gamma ray or hadronic cosmic ray)
%can be established, along with its energy and arrival direction.

VERITAS observations are made at night, under clear skies.  The
telescopes are operated in concert and track gamma-ray source
candidates as they move in azimuth and elevation. The standard
observing mode consists of an exposure of $15-30\U{mins}$, with the
target offset from the center of the field-of-view by $0.5\degree$,
sequentially towards the North, South, East and West. The telescope
positioning is accurate to a few arcminutes, and offline corrections
using CCD star trackers determine individual telescope pointing
directions to better than $20\U{arcsec}$.

%\begin{figure}[h]
%\center \includegraphics[width=0.5\textwidth]{plots/wobbles.png}
%\caption{An illustration of the location of KIC~8462852 within the
% field-of-view of a VERITAS camera. The small circles indicate PMT
% positions, while the larger dashed circles show the track followed
% by KIC~8462852 for different observing offset directions. The thick
% red lines indicate the location of KIC~8462852 during the
% observations by VERITAS.\label{wobbles}}
%\end{figure}
 
Given the wide field-of-view of the instrument, it is common for
interesting targets to be observed serendipitously.  KIC~8462852
(R.A. $20^{\rm h} 06^{\rm m} 15.46^{\rm s}$, Dec
$+44\degree27'24.6''$) lies $1.07\degree$ from 0FGL J2001.0+4352
(R.A. $20^{\rm h} 01^{\rm m} 12.87^{\rm s}$, Dec
$+43\degree52'52.8''$), a gamma-ray source associated with an active
galactic nucleus \citep{2014AnA...572A.121A}. The VERITAS archive
contains observations of this region taken between 2009 and 2015, with
KIC~8462852 offset from the center of the field-of-view by
$0.87\degree$, $1.37\degree$, $0.67\degree$ and $1.48\degree$ for the
N, S, E and W offsets, respectively. Table~\ref{data} lists the
VERITAS observations of KIC~8462852. The total exposure, after
selecting only good-weather runs with no major hardware problems, is
526 minutes.

%Figure~\ref{wobbles} illustrates the
%positions of KIC~8462852 within the VERITAS field-of-view, while

\begin{deluxetable}{cccccc}
\tabletypesize{\scriptsize}
%\rotate
\tablecaption{VERITAS observations of KIC~8462852\label{data}}
\tablewidth{0pt}
\tablehead{
\colhead{MJD} & \colhead{Start} &
\colhead{End} & \colhead{Offset} & \colhead{Elevation} &
\colhead{Comments} \\

\colhead{ } & \colhead{(UTC)} &
\colhead{(UTC)} & \colhead{Direction} & \colhead{Midpoint} &
\colhead{} 
}
\startdata
 55143 & 1:40 & 2:00 & N &  $67.4\degree$    & \\
 55143 & 2:02 & 2:22 & S &   $64.2\degree$   & \\
 55144 & 1:58 & 2:18 & W & $ 63.6\degree$  & \\
 55145 & 1:52 & 2:12 & N &  $64.1\degree$   & \\
 55146 & 1:37 & 1:57 & S &  $66.5\degree$   & \\
 55146 & 1:58 & 2:18 & E &  $63.2\degree$   & \\
 55151 & 1:35 & 1:55 & W &  $62.8\degree$    & \\
 55151 & 1:56 & 2:16 & N & $59.5\degree$   & \\
 55152 & 1:45 & 2:02 & S &  $61.3\degree$   & \\
 55152 & 2:05 & 2:26 & E &  $57.7\degree$   & \\
 55326 & 10:02 & 10:22 & E & $63.0\degree$ & 3 telescopes\\
 55326 & 10:23 & 10:43 & W & $67.4\degree$ & 3 telescopes\\
 55326 & 10:44 & 11:12 & N & $70.6\degree$ & 3 telescopes\\
 55356 & 10:06 & 10:26 & S &  $78.1\degree$   & \\
 55357 & 10:33 & 10:53 & S &  $76.4\degree$   & \\
 55381 & 6:36 & 6:56 & N &  $65.1\degree$   & \\
 55381 & 6:57 & 7:17 & S &  $68.9\degree$  & \\
 56091 & 9:15 & 9:35 & N &  $75.8\degree$   & \\
 56091 & 9:36 & 9:56 & S &   $77.9\degree$  & \\
 \hline
 \hline
 56203 & 4:45 & 5:05 & S &  $60.2\degree$   & \\
 56943 & 2:08 & 2:23 & N &  $76.8\degree$   & \\
 56953 & 1:59 & 2:14 & S &  $75.0\degree$   & \\
 56974 & 1:44 & 1:59 & E &  $64.8\degree$   & \\
 56981 & 1:38 & 1:53 & W &  $60.1\degree$   & \\
 57162 & 9:55 & 10:10 & N &  $68.0\degree$   & \\
 57189 & 8:49 & 9:04 & S &   $74.5\degree$  & \\
 57283 & 4:21 & 4:36 & N &   $73.8\degree$  & \\
 57297 & 2:37 & 2:52 & N &   $77.2\degree$  & \\
\enddata
\tablecomments{The horizontal break indicates when the telescope PMTs
  were upgraded. All four telescopes were operating, unless otherwise stated.}

%\tablenotetext{a}{Sample footnote for table~\ref{tbl-1} that was generated
%with the deluxetable environment}
%\tablenotetext{b}{Another sample footnote for table~\ref{tbl-1}}

%48062 & 300.35267, 44.398657 N 0.8672857
%48063 & 300.35267, 43.39866 S 1.3715912
%48085 & 299.65881, 43.896567 W 1.4770162
%48126 & N
%48174 & S
%48175 & 301.04662, 43.896584 E 0.6721455
%48288 &
%48289 &
%48326 &
%48327 &
%51272 &
%51273 &
%51274 &
%51620 &
%51640 &
%51857 &
%51858 &
%63111 &
%63112 &
%63732 &
%74540 &
%74763 &
%75062 &
%75229 &
%77758 &
%78112 &
%78304 &
%78494 &

%% Text for table notes should follow after the \enddata but before
%% the \end{deluxetable}. Make sure there is at least one \tablenotemark
%% in the table for each \tablenotetext.
%\tablecomments{Table \ref{tbl-1} is published in its entirety in the 
%electronic edition of the {\it Astrophysical Journal}.  A portion is 
%shown here for guidance regarding its form and content.}
%\tablenotetext{a}{Sample footnote for table~\ref{tbl-1} that was generated
%with the deluxetable environment}
%\tablenotetext{b}{Another sample footnote for table~\ref{tbl-1}}
\end{deluxetable}

\section{Analysis and Results}
Data calibration and image pre-processing follow the same procedures
used for gamma-ray observations with VERITAS
\citep{2006APh....25..391H}. The PMT responses are first flat-fielded
using a pulsed LED source \citep{2010NIMPA.612..278H}. For each event,
PMTs containing significant signal are identified, and the resulting
images are parametized with an ellipse. The image properties; its
root-mean-square {\it length} and {\it width}, and its orientation and
intensity \citep{1985ICRC....3..445H},  are then used to classify the
events and, in the case of gamma-ray and cosmic ray images, to
reconstruct the properties of the cascade progenitor.

\begin{figure}
\begin{center}
%\epsscale{.80}
\includegraphics[width=0.5\textwidth]{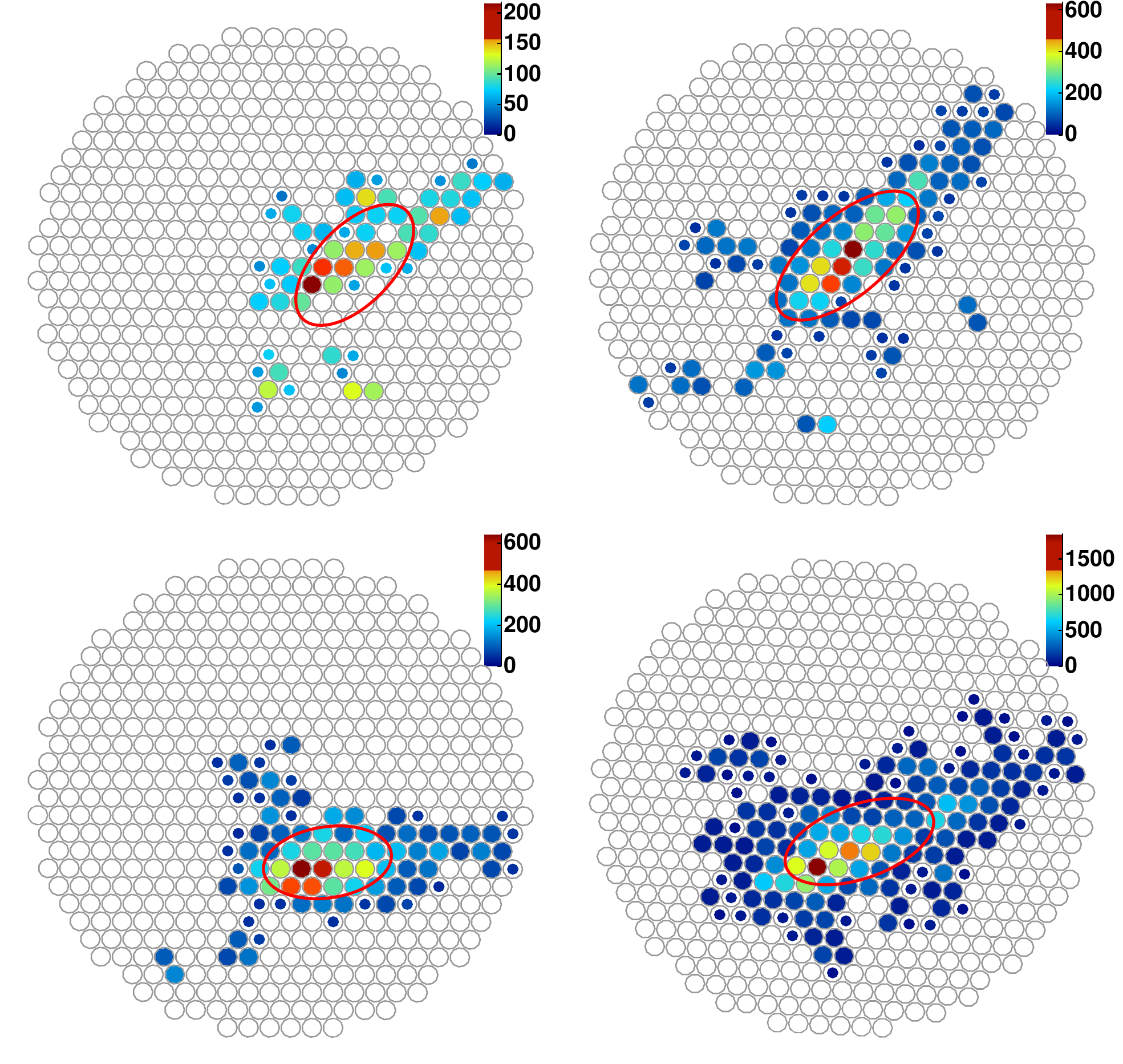}
%\plotone{plots/cosray_large.png}
\vspace{0.25cm}
-----------------------------------------------------------------
\vspace{0.25cm}
\includegraphics[width=0.5\textwidth]{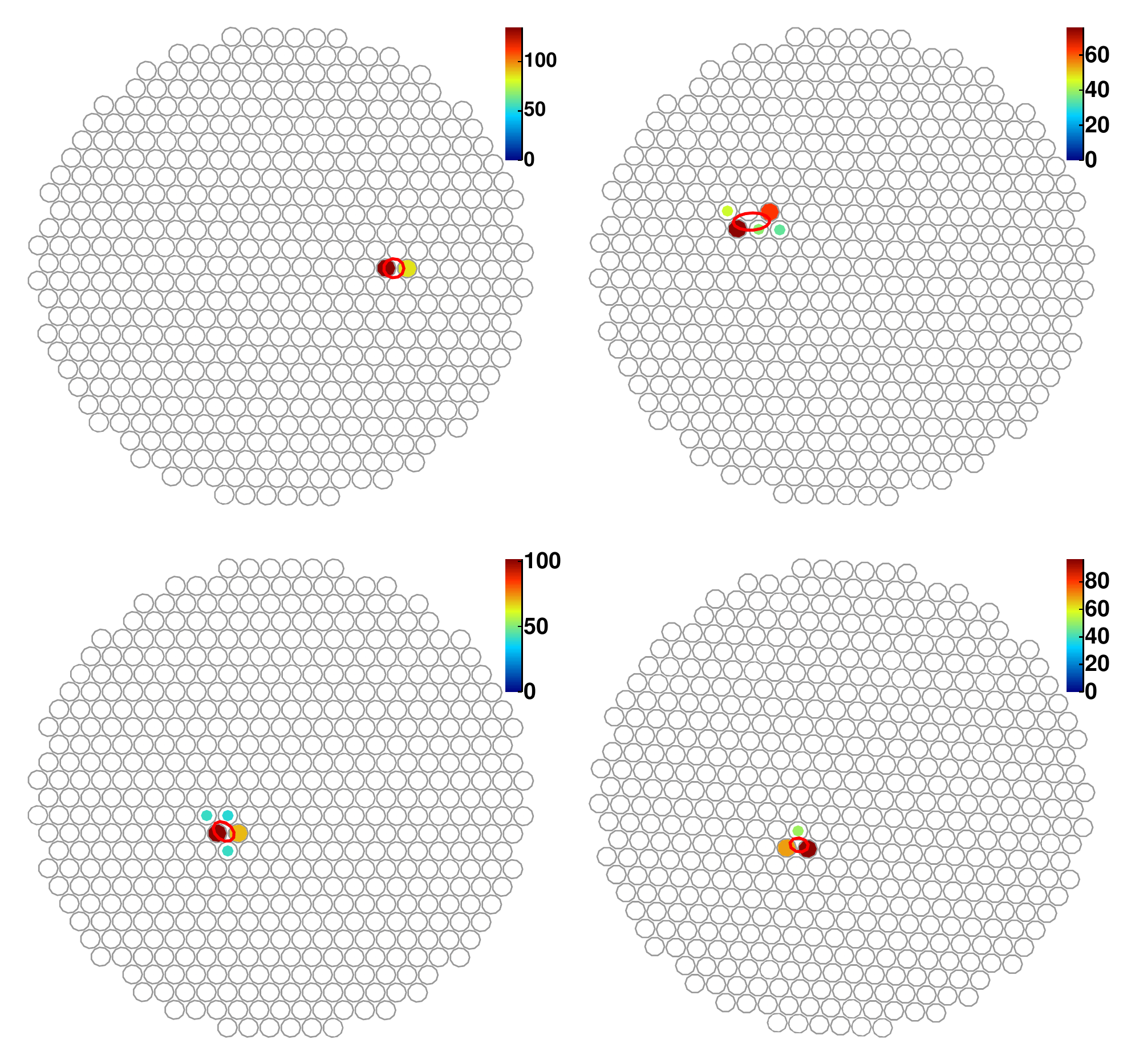}
%\plotone{plots/cosray_small.png}
\end{center}
\caption{Cosmic ray events in the four VERITAS multi-pixel PMT
  cameras. The diameter of the field-of-view is $3.5\degree$, and the
  color scale indicates the intensity of Cherenkov light in each PMT
  in FADC digital counts (where 5.3 digital counts corresponds to one
  photo-electron). The top panel shows an event initiated by a
  high-energy cosmic ray, with large angular extent. The bottom panel
  shows a fainter, lower-energy cosmic ray event, which is still
  easily distinguished from a distant point source due to the large
  parallactic displacement between the image locations in each
  camera. \label{cosrays}}

\end{figure}

In this analysis, we are searching for evidence of pulsed emission
from a distant optical beacon. The resulting images would have a number of
characteristics which make them simple to identify. Specifically:
\begin{itemize}
\item{they appear in the same place in all four telescope cameras}
\item{they have the same intensity in each telescope, and}
\item{they are point-like: that is, they have the  same morphology as
    the telescope optical point-spread function.}
\end{itemize}
These criteria alone are sufficient to remove essentially all of the
background cosmic ray events from the analysis. Cosmic ray air showers
produce the majority of their Cherenkov light below an altitude of
$20\U{km}$, and so the images recorded in detectors separated by
$100\U{m}$ show significant parallactic displacement
($>0.3\degree$). Image centroid locations can be measured with a
precision typically an order of magnitude better than this. Cosmic ray
showers develop and grow longitudinally in the atmosphere, producing
Cherenkov light over a typical length of several kilometers and
resulting in an image with a large angular extent
(Fig.~\ref{cosrays}). Furthermore, an optical beacon originating from
KIC~8462852 would produce images with centroid locations consistent
with the location of the star in the telescope cameras. The most
convincing signal would be given by multiple occurrences of such
images at different times, each matching the star's position.

One concern is whether such images would successfully trigger the
telescope readout, which requires at least three adjacent PMTs in each
telescope camera to have signals exceeding their discriminator
thresholds. The optical point spread function (PSF) at the center of
the field-of-view has a typical full width at half maximum of
$0.06\degree$ (or $0.09\degree$ at the 68\% containment radius), with
almost all of the light in the image of a point source contained
within the diameter of a single PMT ($0.15\degree$). However, as a
natural result of the alt-azimuth design of the telescopes, a
celestial source moves across the field-of-view and spends a
reasonable fraction of time at the interface between two or three PMT
pixels. Furthermore, the optical PSF degrades significantly off the
optical axis, blurring and distorting the image of a point source more
widely near the outer edge of the camera (by approximately 50\% at
$1\degree$ off-axis). This increases the probability that a point-like
image will generate a trigger.

The temporal characteristics of the light pulse also play an important
role in determining the trigger efficiency. The PMT signal path is
AC-coupled in the telescope camera prior to pre-amplification, with a
lower cut-off frequency of approximately $100\U{kHz}$.
%with a time constant of approximately $50\U{\mu s}$ 
A steady, or slowly varying, signal will not generate a trigger. There
is a caveat to this, in that a steady optical source such as a bright
star in the field-of-view will generate high frequency Poisson
noise fluctuations in the PMT signals, increasing the probability of
accidentally crossing the discriminator thresholds. KIC~8462852 itself
has a V-band magnitude of 12.01 \citep{2000AnA...355L..27H}, which is
too faint to produce any measurable increase in the noise fluctuations
of the PMT signal. An optical pulse with a duration much longer than
the AC-coupling time constant, but with a risetime significantly less
than $10\U{\mu s}$, will also trigger the cameras.

Based on these criteria, we have processed all of the data to search
for optical pulses associated with KIC~8462852. For every recorded
event we have applied selection cuts based on the recorded images. We
retained events in which all images had ellipse {\it lengths} and {\it
  widths} less than $0.1125\degree$, equivalent to the {\it length} of
an image with three aligned PMTs each containing exactly the same
signal intensity. We also require that at least three telescopes
record an image, and that no two image centroid locations within the
same event are separated by more than $0.15\degree$, equivalent to the
spacing between adjacent PMTs.

\begin{figure*}
 \includegraphics[width=0.5\textwidth]{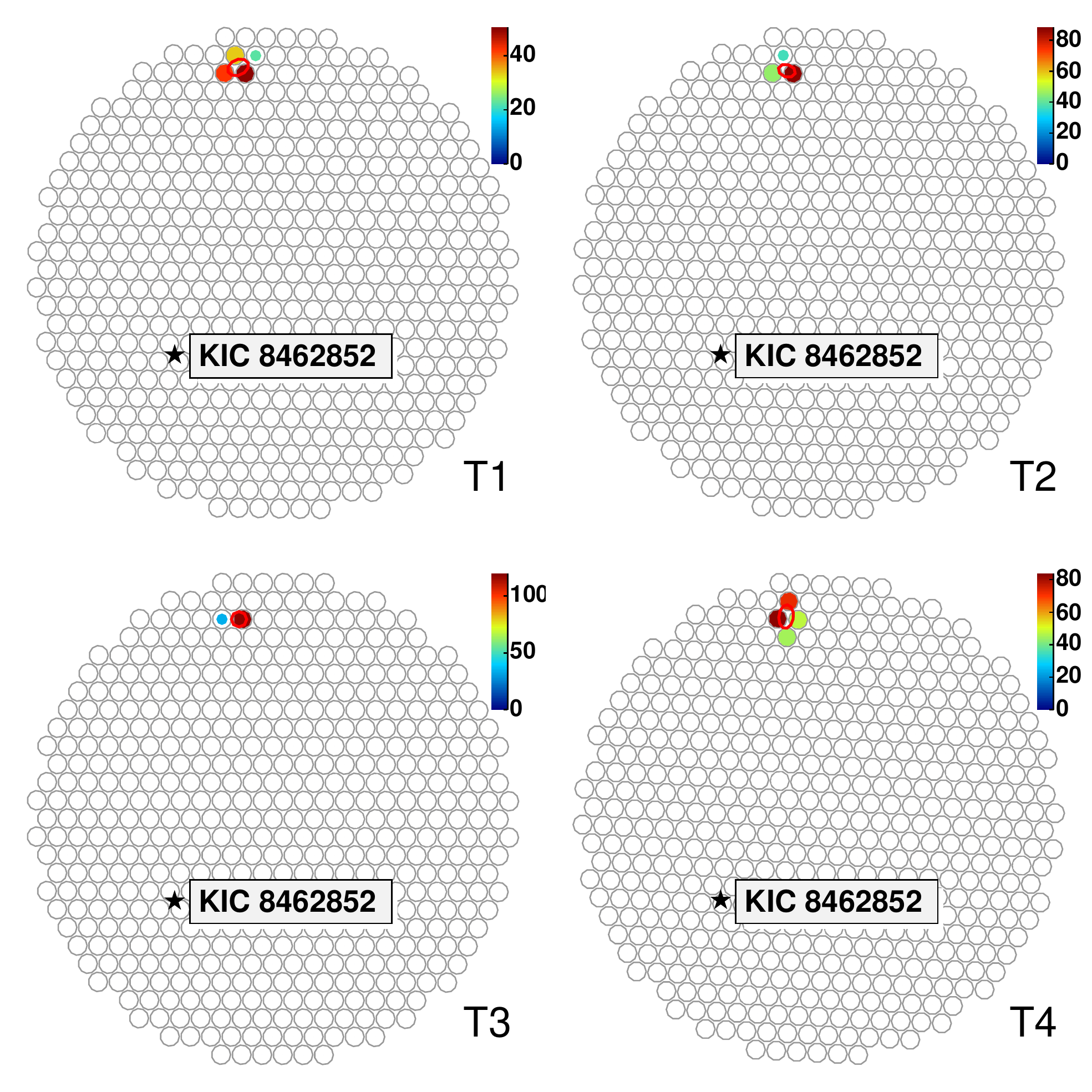}\hspace{0.0\textwidth}\includegraphics[width=0.5\textwidth]{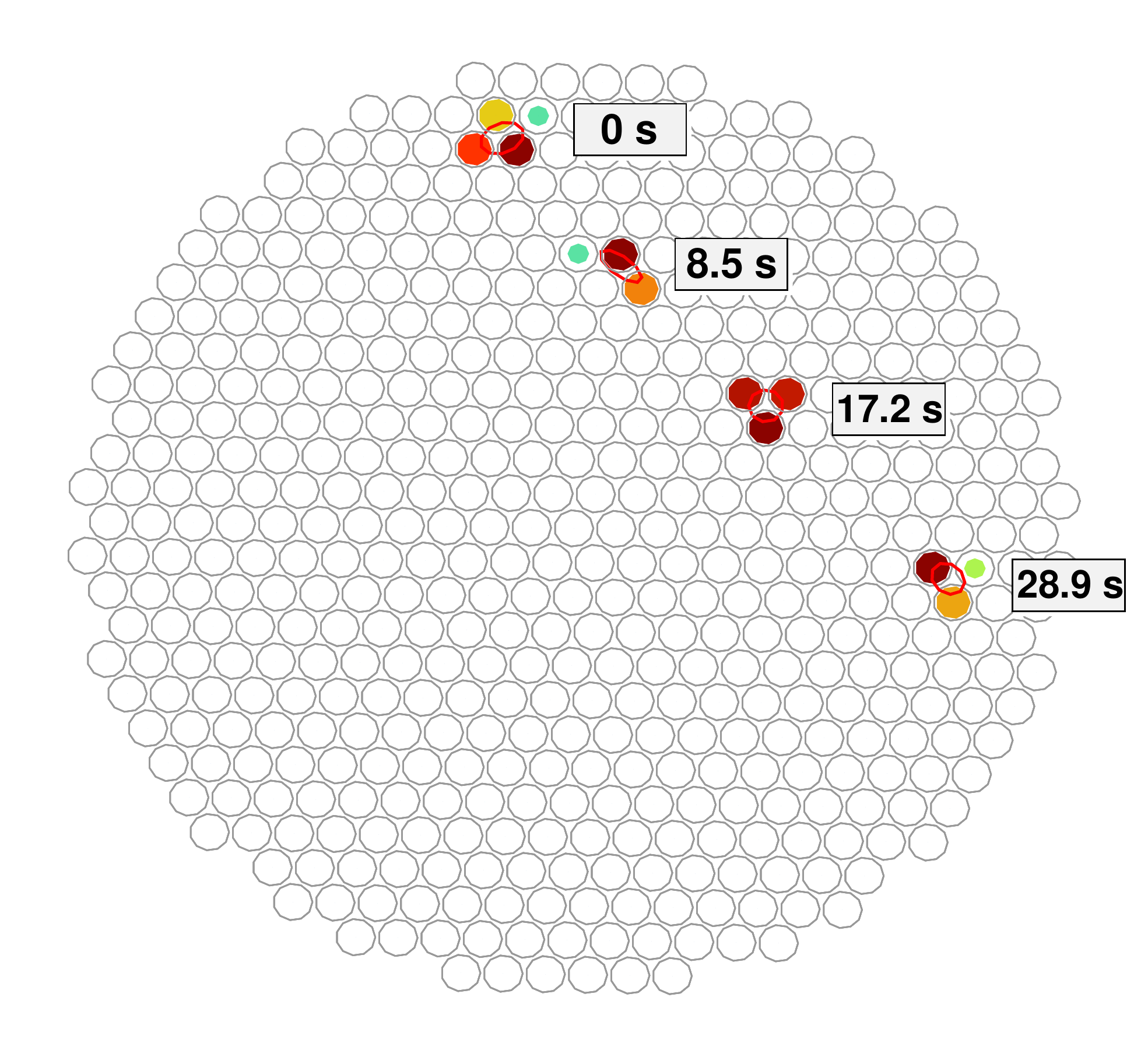}
 \caption{Point-like events generated by an object moving across the
   field-of-view of VERITAS over the course of 28.9 seconds on MJD
   57283. {\bf Left: } A single event viewed by all four
   telescopes. {\bf Right:} A subset of the eight recorded events
   illustrating the motion of the image across the camera of a single
   telescope.}\label{track}
\end{figure*}
 
Inspection of the surviving events reveals that the majority resemble
faint cosmic ray events, or noise fluctuations, typically with low
intensity images. A four-telescope image multiplicity requirement, or
a minimum image intensity cut, efficiently removes these. These
additional criteria, together with a requirement for uniform intensity
across all images, could be be applied in future analyses requiring
stricter background rejection, such as searches without a predefined
source candidate location.  

Interestingly, some point-like optical pulse candidates are
observed. In particular, during a 15-minute exposure on 2015 September
18 (MJD 57283), two different classes of point source optical pulses
are detected. One of these consists of around 300 consecutive events,
which move in a straight line across the field-of-view, crossing
$3.4\degree$ in 0.11 seconds. The images are not co-located in the
field-of-view, displaying a maximum parallactic displacement of
$0.1\degree$, observed between two telescopes separated by
$150\U{m}$. This implies an altitude of emission of $86\U{km}$ and a
velocity of roughly $50\U{km}\UU{s}{-1}$, consistent with the typical
properties of the light from a meteoroid passing through the Earth's
thermosphere. We note that the precursor to VERITAS, the original
Whipple $10\U{m}$ gamma-ray telescope, was used to provide the first
optical observations of meteors down to a limiting photographic
magnitude of +12 in the 1970s \citep{1980MNRAS.193..645C}.  Another
class of optical pulses, an example of which is seen in the same
exposure, are more difficult to explain. These consist of eight
events, recorded with the telescopes pointing towards $323.7\degree$
azimuth and $73.3\degree$ elevation. The events again move in a
straight line across the field-of-view, crossing $2.2\degree$ in 28.9
seconds, from 04:33:01 -- 04:33:30 UT (Fig.~\ref{track}). There is no
measurable parallactic displacement in this case, implying an altitude
of at least $200\U{km}$. The pulse duration is longer than the FADC
read-out window length of $32\U{ns}$. We consider that the most likely
explanation is that these events are due to light reflected from a
satellite; the measured angular velocity ($4.6'\UU{s}{-1}$) is
consistent with an object in orbit at an altitude of a few thousand
kilometers. Events such as these, which form tracks, are easily
distinguished from a true celestial source, simply by their motion
across the field-of-view. They also serve as a useful proof of
principle, demonstrating that the array does trigger on point-like
optical sources at large distance, and highlighting the advantages of
a widely separated telescope array with imaging capabilities and a
large field-of-view in avoiding misclassification.

After applying the selection cuts, and removing any event sequences
which form tracks in the cameras, only 28 of the initial 7036970
events remain -- 1 in every 251320 events.  None of these have images
consistent with the location of KIC~8462852. We conclude that there is
no evidence in the VERITAS observations for optical pulses originating
from this system.

\section{Discussion}

\citet{2009AsBio...9..345H} have already outlined the difficulty in
forming meaningful constraints from optical SETI observations. In
particular, aside from the very obvious reasons for the lack of a
detection (i.e. there is very likely no extraterrestrial civilization
at KIC~8462852 directing laser pulses towards us) the potentially
transient nature of any signal implies that the same observations
could be performed at a different time, with a different
result. Nevertheless, with nine hours of observations distributed over
six years, the VERITAS archive represents a unique dataset.

An exact estimate of the sensitivity of the search is also difficult,
since the minimum detectable optical pulse intensity depends strongly
upon the emission wavelength, the duration and temporal profile of the
pulse, and the exact source location within the field-of-view when the
pulse occurs -- all of which are unknown. Taking representative values
(20\% photon detection efficiency, 85\% mirror reflectivity, and a
conservative minimum image intensity of 100 digital counts,
corresponding to 18.8 photo-electrons) gives a minimum optical pulse
intensity of $0.94\U{ph}\UU{m}{-2}$ arriving at the telescope within
the $12\U{ns}$ pulse integration window used for this analysis. The
required energy of the transmitted pulse, estimated using the method
of \citet{2004ApJ...613.1270H}, is $0.3\U{MJ}$, or roughly equivalent
to a $B-$band 6.4 mag star, but with just a few nanoseconds
duration. This compares favorably with earlier searches with optical
telescopes, which have sensitivities of typically
$60-100\U{ph}\UU{m}{-2}$, and with observations by the STACEE
heliostat array telescope, which provided a sensitivity of
$10\U{ph}\UU{m}{-2}$ \citep{2009AsBio...9..345H}. The recent optical
SETI observations of KIC~8462852 by \citet{2015arXiv151202388S} report
a sensitivity to periodic signals of $67\U{ph}\UU{m}{-2}$.

%Over the coming months, we
%plan to test the array trigger and read-out response using
%artificially generated point-like optical flashes of varying duration,
%intensity and wavelength, which will provide more detailed information
%on the sensitivity to optical pulses with different characteristics.

\begin{figure}[h]
\center \includegraphics[width=0.5\textwidth]{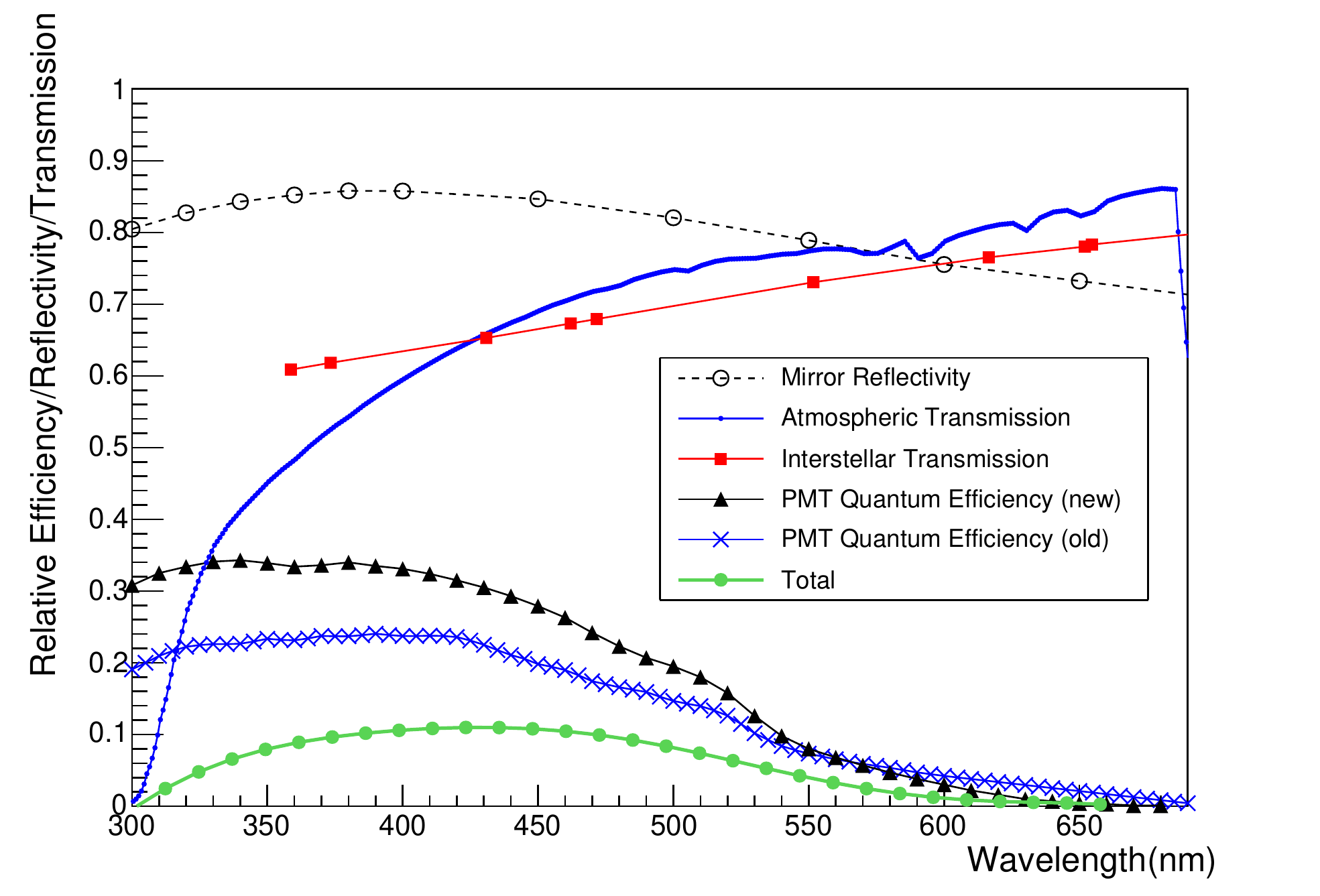}
\caption{The wavelength-dependent efficiency for detection of an
  optical beacon by VERITAS, illustrating the effects of interstellar
  reddening, the Earth's atmosphere (generated using MODTRAN
  \citep{2006SPIE}), the mirror reflectivity (VERITAS measurement),
  and the PMT quantum efficiency both before and after the PMT upgrade
  in 2012 (VERITAS measurement). The ``Total'' curve multiplies these
  effects together (assuming the post-2012 quantum efficiency), giving
  a peak efficiency for detection at $420\U{nm}$. \label{efficiency}}
\end{figure}

We can also estimate the wavelength dependence of the sensitivity,
taking into account the effects of interstellar absorption,
transmission through the Earth's atmosphere, the reflectivity of the
telescope mirrors and the quantum efficiency of the photo-detectors.
\citet{2015arXiv150903622B} calculate a distance to KIC~8462852 of
$454\U{pc}$, based upon the interstellar reddening of the spectral
energy distribution ($E(B-V)=0.11\pm0.03\U{mag}$), which corresponds
to a $V$-band extinction of $A_V=0.341\U{mag}$, or a transmission
efficiency of 73\%. We normalize wavelength-dependent extinction
values \citep{2011ApJ...737..103S} to this measurement, as shown in
Figure~\ref{efficiency}. The overall wavelength dependence of the
sensitivity is given by the ``Total'' curve.

Perhaps the most important conclusion of this work is that modern IACT
arrays are effective tools to search for faint optical transients,
with durations as brief as a few nanoseconds. The observations are
complementary to, and have no impact on, the scientific program in the
gamma-ray domain. Any artificial backgrounds can be easily removed,
using the simple analysis procedures outlined here. The search for
short duration optical transients was highlighted as a fruitful area
of research almost half a century ago
(e.g. \citet{1970QJRAS..11..443B} and \citet{1984ApJ...283..887S}) and
has developed into an important field of astronomy over the past
decade, with many different scientific goals
(e.g. \citet{2012IAUS..285....9S} lists gravitational microlensing,
asteroid detection, stellar variability, extragalactic transients and
cosmology using supernovae).  Dedicated robotic arrays of astrographs,
or reflecting telescopes with Schmidt optics, monitor the skies
nightly to explore the time domain (e.g. the Catalina Real-Time
Transient Survey, the Hungarian Automated Telescope Network, the
Palomar Transient Factory, etc.) while, in the near future, the Large
Synoptic Survey Telescope will map the entire visible sky every few
nights with pairs of $15\U{s}$ exposures, separated by 15 --
$60\U{minutes}$ and reaching a $5\sigma$ sensitivity limit of
$24.5\U{mag}$ \citep{2008arXiv0805.2366I}. Extending this search to
nanosecond timescales opens up a new area of parameter space. While
short-duration gamma-ray bursts have now been studied for decades, the
surprising recent discovery of millisecond-timescale fast radio bursts
\citep{2007Sci...318..777L}, and nanosecond radio pulses from the Crab
pulsar \citep{2003Natur.422..141H} demonstrates that such rapid
phenomena may exist elsewhere in the electromagnetic spectrum. Further
motivation is provided by \citet{2013ApJ...774..142B}, who has
discovered evidence for rapid periodic modulations in the optical
spectra of galaxies, while \citet{2014MNRAS.445.1858L} has noted the
potential of IACTs for rapid photometry of bright sources,
particularly in the context of observing stellar occultations by small
objects in the outer solar system.

VERITAS has been operating since 2007 and records over $1000\U{hours}$
of observations per year. Together with the H.E.S.S. array in Namibia
and the MAGIC system in La Palma, approximately $30,000\U{hours}$ of
archived IACT data exist and a substantial fraction of the entire sky
has been observed. A search of these archives for optical transients
seems worthwhile. Beyond the current generation of instruments lies
the planned Cherenkov Telescope Array (CTA), which will consist of
northern and southern hemisphere telescope arrays on a much larger
scale, totalling over 100 telescopes \citep{2013APh....43....3A}. CTA
will provide an enormous increase in mirror area and telescope
multiplicity, with the potential to greatly enhance searches for
optical transients. This additional application of the facility can
inform design decisions regarding the telescope optics and trigger
systems.

\acknowledgments

This research is supported by grants from the U.S. Department of
Energy Office of Science, the U.S. National Science Foundation and the
Smithsonian Institution, and by NSERC in Canada. We acknowledge the
excellent work of the technical support staff at the Fred Lawrence
Whipple Observatory and at the collaborating institutions in the
construction and operation of the instrument. The VERITAS
Collaboration is grateful to Trevor Weekes for his seminal
contributions and leadership in the field of VHE gamma-ray
astrophysics, and for his interest in the wider applications of IACTs,
which made this study possible.

{\it Facilities:} \facility{VERITAS}

\end{document}